\documentclass[a4paper]{scrartcl}

\usepackage{amsmath}
\usepackage{amsfonts}
\usepackage{amssymb}
\usepackage{color}
\usepackage[pdftex,colorlinks=true]{hyperref}
\usepackage{thumbpdf}
\usepackage{listings}
\usepackage{biblatex}
\usepackage{booktabs}
\usepackage{enumitem}
\bibliography{funcall.bib}

\newcommand{\mytitle}{Function call overhead benchmarks}
\newcommand{\mysubtitle}{with MATLAB, Octave, Python, Cython and C}
\newcommand{\myauthor}{Andr\'{e} Gaul}
\hypersetup{
	pdftitle={\mytitle \mysubtitle},
	pdfauthor={\myauthor},
	pdfkeywords={function call, loop, overhead, MATLAB, Octave, Python,
        Cython, C}
}

\title{\mytitle}
\subtitle{\mysubtitle}
\author{\myauthor\footnote{
\href{mailto:gaul@math.tu-berlin.de}{\texttt{gaul@math.tu-berlin.de}},
\url{http://www.math.tu-berlin.de/?78347}.
This work is licensed under the creative commons license CC-BY-3.0. All
used source code files are published under GPL 3.0 in the git repository 
\url{https://bitbucket.org/andrenarchy/funcall}.
}}
\date{\today}

\definecolor{dkgreen}{rgb}{0,0.6,0}
\definecolor{gray}{rgb}{0.5,0.5,0.5}
\definecolor{mauve}{rgb}{0.58,0,0.82}
\begin{document}
    \maketitle

    \section{Background}
    \label{sec:background}

    In many applications a function has to be called very often inside a loop.
    One such application in numerical analysis is the finite element method
    (FEM) for the approximate solution of a partial differential equation (PDE).
    For example we would like to approximate the solution
    $u:\Omega\rightarrow\mathbb{R}$ of the PDE
    \begin{align*}
        -\nabla \cdot (a\nabla u) + b\cdot \nabla u + cu &= f \qquad \text{in}
        \quad \Omega \\
        u &= g \qquad \text{on} \quad \partial \Omega
    \end{align*}
    where $\Omega\subseteq\mathbb{R}^d$ is a $d$-dimensional domain with
    boundary $\partial\Omega$ and
    $a,c,f:\Omega\rightarrow\mathbb{R}$, $b:\Omega\rightarrow\mathbb{R}^d$ and
    $g:\partial\Omega\rightarrow\mathbb{R}$ are given functions with special
    properties that will not be discussed here. In FEM a domain
    is discretized into a mesh by splitting the domain into ``simple''
    geometric shapes (intervals, triangles, tetrahedrons, \ldots).  Along with
    special functions (usually piecewise polynomials) these shapes are called
    \emph{elements}. A system of linear algebraic equations $Ax=b$ is obtained
    by computing integrals on each element and sorting them into a large (and
    usually sparse) ``system'' matrix $A$ and right hand side (RHS) $b$.

    From a computational point of view this can be achieved by writing a
    function \texttt{getElementIntegrals(...)} that computes all
    necessary integrals on one element and is then called within a loop for each
    element of the mesh. A corresponding Python code could look like this:
    \begin{lstlisting}
for el in mesh:
    elMat, elRHS = getElementIntegrals(el,a,b,c,f,g)
    # process entries of elMat and elRHS by sorting them 
    # into a "big" matrix and right hand side.
    \end{lstlisting}
    Now imagine that the mesh is very fine, i.e. the number of elements in the
    mesh is large. For example a uniform tetrahedral discretization of the unit
    cube $[0,1]^3$ with grid size $h=1/n$ ($n\in\mathbb{N})$ results in a mesh
    consisting of $6n^3$ tetrahedral elements. The function
    \texttt{getElementIntegrals} is thus called $6n^3$ times. 
    
    Especially in interpreted programming languages like MATLAB, Octave or Python
    a function call may be very time-consuming. By \emph{function call} we mean
    the setup needed for the function to start executing the actual function
    code in the function's body and cleaning it up afterwards. This includes
    possible copying of memory and dynamic type checking for the parameters
    passed to and returned from the function.
    
    In the above setting the functions \texttt{a}, \texttt{b}, \texttt{c},
    \texttt{f}, \texttt{g} and even \texttt{getElementIntegrals} can often be
    evaluated in a fast way. However, the function call itself may exhibit an
    overhead that consumes far more time than the actual function code. In this
    report we present results of function call benchmarks for programming
    languages or interpreters often used in numerical analysis: MATLAB/Octave,
    Python and C. While C is a compiled language and
    optimizations are possible even on a very low level, MATLAB and the free software
    alternative Octave are interpreted languages and mainly draw on automated
    optimization and low-level improvements are usually only possible by
    switching to plain C, C++ or Fortran with so-called mex-files. In contrast,
    in the interpreted language Python time-critical parts can be compiled with
    Cython -- the C-Extensions for Python~\cite{Cython}. Cython's syntax is very
    similar to Python's and introduces static typing as well as the ability to
    call C code easily. 
    
    Here we present a benchmark that helps to identify and quantify optimization
    potentials with respect to time consumption caused by function calls in the
    mentioned languages.  Section~\ref{sec:setup} describes the setup of the
    benchmark and Section~\ref{sec:results} presents and discusses the results.
    
    \section{Benchmark setup}
    \label{sec:setup}

    The situation outlined in Section~\ref{sec:background} can be boiled down
    to a function that is called in a loop very often. For benchmarking the
    overhead of the function call itself it is reasonable to make the function
    body as simple as possible. Therefore, we will use a function that
    accepts one double precision floating-point number and returns its square.
    We ran separate tests for several types of function definitions that are available in
    the used programming languages. The different options are enumerated for
    later reference.
    \begin{description}
        \item[MATLAB and Octave]~
            \begin{enumerate}[label=Option \arabic{*}., ref=\arabic{*}]
                \item \label{enu:mat}
                    \begin{enumerate}[label=(\alph{*}), ref=(\alph{*})]
                        \item \label{enu:mat:ext} 
                            The called function defined in an external .m-file:
                            \begin{lstlisting}[language=MATLAB]
function result = fun_external(a)
    result = a*a;
end
                            \end{lstlisting}
                            The loop is defined in a separate file:
                            \begin{lstlisting}[language=MATLAB]
function result = loop_external(n)
    result = zeros(n,1);
    for i=1:n
        result(i) = fun_external(i);
    end
end
                            \end{lstlisting}
                        \item \label{enu:mat:same}
                            Both functions from~\ref{enu:mat:ext} are placed in
                            the same file consecutively. 
                        \item \label{enu:mat:nest}
                            A nested function definition in the body of the
                            calling function is used, that is the called function is
                            placed \emph{inside} the body of the loop function:
                            \begin{lstlisting}[language=MATLAB]
function result = loop_nested(n)
    result = zeros(n,1);
    for i=1:n
        result(i) = fun_nested(i);
    end

    function ret = fun_nested(a)
        ret = a*a;
    end
end
                            \end{lstlisting}
                        \item \label{enu:mat:anon}
                            Anonymous function definition in the body of the
                            calling function:
                            \begin{lstlisting}[language=MATLAB]
fun_anonymous = @(a) (a*a);
                            \end{lstlisting}
                    \end{enumerate}
            \end{enumerate}
        \item[Python] ~
            
            Python is an interpreted language but can be tuned by
            writing time-critical parts in Cython~\cite{Cython}. With Cython one
            can blend Python code and C code easily. We take a closer look at
            the following options for implementing the loop and the called
            function:
            \begin{enumerate}[label=Option \arabic{*}., ref=\arabic{*}]
                \item \label{enu:py:file}
                    The called function can be implemented 
                    \begin{enumerate}[label=(\alph{*}), ref=(\alph{*})]
                        \item \label{enu:py:file:same}
                            together with the loop function in the same
                            .py-file or
                        \item \label{enu:py:file:imp}
                            in a separate .py-file and \texttt{import}ed in the
                            .py-file implementing the loop.
                    \end{enumerate}
                \item \label{enu:py:loop} 
                    The loop can be implemented with a numpy array~\cite{SciPy} in 
                    \begin{enumerate}[label=(\alph{*}), ref=(\alph{*})]
                        \item \label{enu:py:loop:std} 
                            plain Python:
                            \begin{lstlisting}
def loop(n):
    result = numpy.empty(n)
    for i in xrange(0,n):
        result[i] = fun_samefile(i+1)
    return result
                            \end{lstlisting}
                            or in 
                        \item \label{enu:py:loop:cy}
                            Cython where the code still is plain Python code as
                            in~\ref{enu:py:loop:std} or in
                        \item \label{enu:py:loop:cyty} 
                            Cython enriched with static typing:
                            \begin{lstlisting}
def loop(n):
    cdef numpy.ndarray[numpy.double_t] result = \
            numpy.empty(n)
    cdef int i
    for i in xrange(0,n):
        result[i] = fun_samefile(i+1)
    return result
                            \end{lstlisting}
                            Note that the \texttt{result} array and the
                            \texttt{i} variable are now typed which allows
                            Cython to address the elements of the numpy array in
                            the loop efficiently. 
                    \end{enumerate}
                \item \label{enu:py:fun} 
                    Similarly, the called function can be implemented in
                    \begin{enumerate}[label=(\alph{*}), ref=(\alph{*})]
                        \item \label{enu:py:fun:std}
                            plain Python:
                            \begin{lstlisting}
def fun(a):
    return a**2
                            \end{lstlisting}
                            or in 
                        \item \label{enu:py:fun:cy}
                            Cython where the code still is plain Python code as
                            in~\ref{enu:py:fun:std} or in
                        \item \label{enu:py:fun:cyty} 
                            Cython enriched with static typing:
                            \begin{lstlisting}
cpdef double fun(double a):
    return a**2
                            \end{lstlisting}
                            If the function is \texttt{import}ed in the
                            .py-file running the loop then an additional
                            .pxd-file with the corresponding function
                            declaration should be provided. A .pxd-files works
                            like a C header file and in our case simply contains
                            the line
                            \begin{lstlisting}
cpdef double fun(double a)
                            \end{lstlisting}
                    \end{enumerate}
            \end{enumerate}
            Several combinations are not possible and are thus omitted. For
            example,
            option~\ref{enu:py:file}~\ref{enu:py:file:same} with 
            option~\ref{enu:py:loop}~\ref{enu:py:loop:std} and
            option~\ref{enu:py:fun}~\ref{enu:py:fun:cy} are impossible because
            both the loop and the called function are compiled with Cython if
            they are defined in the same file).

        \item[C]~
            \begin{enumerate}[label=Option \arabic{*}., ref=\arabic{*}]
                \item\label{enu:c}
                    \begin{enumerate}[label=(\alph{*}), ref=(\alph{*})]
                        \item \label{enu:c:stat}
                            The function is defined in the same
                            .c-file as the loop and compiled with the options
                            \texttt{-O3 -fomit-frame-pointer}. The function code is
                            \begin{lstlisting}[language=C]
double fun(double a) {
    return a*a;
}
                            \end{lstlisting}
                            while the loop code is
                            \begin{lstlisting}[language=C]
double* loop(int n) {
    double* result = 
            (double*) malloc(sizeof(double)*n);
    for (int i=0; i<n; i++)
        result[i] = fun(i);
    free(result);
    return result;
}
                            \end{lstlisting}
                        \item \label{enu:c:dyn}
                            The function is compiled in a shared library
                            (.so-file) which is then dynamically linked to the
                            compiled loop function. The compiler options are the
                            same as for~\ref{enu:c:stat}.
                    \end{enumerate}
            \end{enumerate}
    \end{description}

    For further details we refer to the source code~\cite{Gau12:funcall}.
    
    \section{Benchmark results}
    \label{sec:results}

    In this section we present results of the benchmark setup described in
    Section~\ref{sec:setup} conducted with the languages/interpreters
    \begin{itemize}
        \item MATLAB 2011b
        \item Octave 3.2.4
        \item Python 2.7.2 
        \item Cython 0.14.1 (C-Extensions for Python)
        \item C with GCC 4.6.1.
    \end{itemize}
    All experiments have been carried out on a Intel Core i5 M540 CPU running at
    2.53~GHz with Ubuntu 11.10. We computed $a^2$ for $a=1,\ldots,10^7$ with all
    possible variations of implementations with the options presented
    in~\ref{sec:setup}. By using this test setup we wish to identify and
    quantify possibilities for optimization with respect to time consumption
    caused by function calls. The experiment was repeated 10 times and the
    arithmetic mean of the measured timings are presented in
    Table~\ref{tab:results}. 
    
    The files used for the experiments are published~\cite{Gau12:funcall} under
    GPL3 so further results can be produced with later versions of the above
    software and on different hardware. 

    \begin{table}[htb!]
        \begin{center}
            \begin{tabular}{rrlccc}
                \toprule
                Rank &Time in s&Language &\multicolumn{3}{c}{Variant}\\
                \cmidrule(r){4-6}
                     &         &            & Option 1                &Option 2                 &Option 3\\
                \midrule
                1  &  0.055 &C              & \ref{enu:c:stat}        & --                      & --                    \\ %&\reftag{enu:c}\\ 
                2  &  0.076 &C              & \ref{enu:c:dyn}         & --                      & --                    \\ %&\reftag{enu:c}\\ 
                3  &  0.077 &Python/Cython  & \ref{enu:py:file:imp}   & \ref{enu:py:loop:cyty}  & \ref{enu:py:fun:cyty} \\ %&\refcusttag{(PyLoopCyTyFunCyTy)}\\
                4  &  0.077 &Python/Cython  & \ref{enu:py:file:same}  & \ref{enu:py:loop:cyty}  & \ref{enu:py:fun:cyty} \\ %&\refcusttag{(PySameLoopCyTyFunCyTy)}\\
                5  &  1.283 &Python/Cython  & \ref{enu:py:file:same}  & \ref{enu:py:loop:cyty}  & \ref{enu:py:fun:cy}   \\ %&\refcusttag{(PySameLoopCyTyFunCy)}\\
                6  &  1.323 &Python/Cython  & \ref{enu:py:file:same}  & \ref{enu:py:loop:cy}    & \ref{enu:py:fun:cyty} \\ %&\refcusttag{(PySameLoopCyFunCyTy)}\\
                7  &  1.337 &Python/Cython  & \ref{enu:py:file:imp}   & \ref{enu:py:loop:cy}    & \ref{enu:py:fun:cyty} \\ %&\refcusttag{(PyLoopCyFunCyTy)}\\
                8  &  1.598 &Python/Cython  & \ref{enu:py:file:imp}   & \ref{enu:py:loop:cyty}  & \ref{enu:py:fun:cy}   \\ %&\refcusttag{(PyLoopCyTyFunCy)}\\
                9  &  2.124 &Python/Cython  & \ref{enu:py:file:same}  & \ref{enu:py:loop:cy}    & \ref{enu:py:fun:cy}   \\ %&\refcusttag{(PySameLoopCyFunCy)}\\
                10 &  2.298 &Python/Cython  & \ref{enu:py:file:imp}   & \ref{enu:py:loop:std}   & \ref{enu:py:fun:cyty} \\ %&\refcusttag{(PyLoopFunCyTy)}\\
                11 &  2.426 &Python/Cython  & \ref{enu:py:file:imp}   & \ref{enu:py:loop:std}   & \ref{enu:py:fun:cy}   \\ %&\refcusttag{(PyLoopFunCy)}\\
                12 &  2.553 &Python/Cython  & \ref{enu:py:file:imp}   & \ref{enu:py:loop:cyty}  & \ref{enu:py:fun:std}  \\ %&\refcusttag{(PyLoopCyTyFun)}\\
                13 &  2.862 &Python/Cython  & \ref{enu:py:file:imp}   & \ref{enu:py:loop:cy}    & \ref{enu:py:fun:cy}   \\ %&\refcusttag{(PyLoopCyFunCy)}\\
                14 &  2.941 &Python         & \ref{enu:py:file:same}  & \ref{enu:py:loop:std}   & \ref{enu:py:fun:std}  \\ %&\refcusttag{(PySameLoopFun)}\\
                15 &  2.973 &MATLAB         & \ref{enu:mat:same}      & --                      & --                    \\ %&\reftag{enu:matsame}\\
                16 &  3.018 &MATLAB         & \ref{enu:mat:ext}       & --                      & --                    \\ %&\reftag{enu:matext}\\
                17 &  3.359 &Python/Cython  & \ref{enu:py:file:imp}   & \ref{enu:py:loop:cy}    & \ref{enu:py:fun:std}  \\ %&\refcusttag{(PyLoopCyFun)}\\
                18 &  3.715 &Python         & \ref{enu:py:file:imp}   & \ref{enu:py:loop:std}   & \ref{enu:py:fun:std}  \\ %&\refcusttag{(PyLoopFun)}\\
                19 &  4.181 &MATLAB         & \ref{enu:mat:nest}      & --                      & --                    \\
                20 &  6.590 &MATLAB         & \ref{enu:mat:anon}      & --                      & --                    \\ %&\reftag{enu:matanon}\\
                21 &112.154 &Octave         & \ref{enu:mat:anon}      & --                      & --                    \\ %&\reftag{enu:matanon}\\
                22 &133.725 &Octave         & \ref{enu:mat:nest}      & --                      & --                    \\ %&\reftag{enu:matext}\\
                23 &138.603 &Octave         & \ref{enu:mat:same}      & --                      & --                    \\ %&\reftag{enu:matsame}\\
                24 &152.452 &Octave         & \ref{enu:mat:ext}       & --                      & --                    \\ %&\reftag{enu:matext}\\
                \bottomrule
            \end{tabular}
        \end{center}
        \caption{Execution time in seconds for $10^7$ function calls. The
        variants are described in Section~\ref{sec:setup}.}
        \label{tab:results}
    \end{table}

    Unsurprisingly, both C implementations with enabled compiler
    optimizations are the fastest implementations in this benchmark. The fact
    that Octave is slower with function calls than MATLAB is also well-known.
    More interesting is the observation that Python and MATLAB approximately
    consume the same amount of time when no optimizations are used in Python.
    However, we can see in the first lines of Table~\ref{tab:results} that
    Python can be tuned with the C-Extensions Cython such that the execution
    time reaches the one of the dynamically linked C implementation, which is
    about 40 times faster than the plain Python or MATLAB implementation. 
    
    The possibility to write performance-critical parts in the Python-like
    Cython syntax is a clear advantage over MATLAB and Octave because currently
    an optimization of function calls in MATLAB can only be achieved by writing
    .mex-files that require a complete rewrite of the code in another language
    and are often hard to handle -- especially if several versions of MATLAB and
    thus the mex-API are used. However, we want to point out that in principle
    the performance of the C variants can be achieved with mex-based
    implementations by calling C code in the mex-files. 
    
    The Cython approach requires less effort since the Cython syntax is very 
    similar to the plain Python syntax. Thus optimizations can be implemented
    easily with Cython where they are needed while maintaining the full
    flexibility of Python.

    \printbibliography

\end{document}